\documentclass[preprint,showpacs,showkeys,aps]{revtex4}
\usepackage{graphicx}
\begin{document}
\def\be{\begin{equation}}
\def\ee{\end{equation}}
\def\bea{\begin{eqnarray}}
\def\eea{\end{eqnarray}}
\title{Exterior and interior metrics with quadrupole moment}
\author{Hernando Quevedo }
\email{quevedo@nucleares.unam.mx}    
\affiliation{
Dipartimento di Fisica, Universit\`a di Roma "La Sapienza", Piazzale Aldo Moro 5, I-00185 Roma, Italy;  
ICRANet, Piazza della Repubblica 10, I-65122 Pescara, Italy.}
\thanks{On sabbatical leave from Instituto de Ciencias Nucleares, Universidad Nacional Aut\'onoma de M\'exico}

\begin{abstract}
We present the Ernst potential and the line element of an exact solution of Einstein's vacuum field equations
that contains as arbitrary parameters the total mass, the angular momentum, and the quadrupole moment of 
a rotating mass distribution. We show that in the limiting case of slowly rotating and slightly deformed 
configuration, there exists a coordinate transformation that relates the exact solution with the 
approximate Hartle solution. It is shown that this approximate solution can be smoothly matched with an
interior perfect fluid solution with physically reasonable properties. This opens the possibility of considering
the quadrupole moment as an additional physical degree of freedom that could be used to search 
for a realistic exact solution, representing both the interior and exterior gravitational field generated by a
self-gravitating axisymmetric distribution 
of mass of perfect fluid in stationary rotation.
\end{abstract}

\maketitle

\section{Introduction}
\label{sec:int}
Astrophysical compact objects are in general not spherically symmetric rotating mass distributions. To describe the corresponding 
gravitational field one can assume axial symmetry, with an axis of symmetry that coincides with the axis of rotation. Then, the 
deviation from spherical symmetry can be described by means of axisymmetric multipole moments. 

In general relativity, vacuum solutions
with multipole moments have been known for a long time. In fact, if we restrict ourselves to solutions with only monopole moment, Birkoff's
theorem guarantees that the only solution is the Schwarzschild metric. Moreover, if rotation is also taken into account, the black hole uniqueness
theorems state that the only asymptotically flat solution with a regular horizon is the Kerr metric \cite{solutions}. The next 
interesting multipole is the quadrupole. In this case, the uniqueness theorems do not apply and it is possible to find a large number
of different vacuum solutions with the same quadrupole. Differences appear only at the level of higher multipoles.
The first static solution with an arbitrary quadrupole was found by Weyl \cite{solutions}, using cylindrical coordinates. 
Later on, Erez and Rosen (ER) discovered a solution with arbitrary quadrupole in prolate spheroidal coordinates which are more convenient 
for the investigation of multipole solutions. Zipoy \cite{zip66} and Voorhees \cite{voor70} 
found a simple transformation which allows to generate static 
solutions from a given one. In particular, the Zipoy-Voorhees (ZV) transformation can be used to generate the simplest solution with quadrupole, 
starting from the Schwarzschild metric. Stationary solutions represent an additional challenge. In fact, the first physically relevant rotating solution was found only in 1963 by Kerr \cite{kerr63}. The Ernst representation of stationary axisymmetric fields 
was an important achievement that allowed to search for the symmetries of the field equations upon which modern solution generating techniques are
based. 

In this work, 
we will limit ourselves to the study of a particular stationary axisymmetric vacuum solution which was derived by Quevedo and Mashhoon (QM) 
in \cite{qm85,quev86} as  a generalization of the ER metric \cite{erro59}.  The QM solution contains in general an infinite 
number of gravitational and electromagnetic multipole moments. Here, however, we neglect the electromagnetic field and
focus on the contribution of the gravitational quadrupole only. 
The main goal of the present work is to show that the QM solution can be used to describe the exterior gravitational field of rotating
compact objects. First, we will see that the set of independent and arbitrary parameters entering the metric determines the total mass, angular 
momentum and mass quadrupole moment of the source. Although higher multipole moments are present, they all can be expressed in terms of 
the independent lower multipoles. This result is obtained by using the invariant definition of relativistic multipole moments proposed by Geroch and
Hansen \cite{ger,hans}.  The spacetime turns out to be asymptotically flat and and free of singularities outside a region which can be 
``covered" by an interior perfect fluid solution. This last property, however, is shown only in the limiting case of 
a slightly deformed body with uniform and slow rotation. 

There exists in the literature a reasonable number of interior spherically symmetric solutions  which can be matched with the exterior Schwarzschild metric. Nevertheless, a major problem of classical general relativity consists in finding a physically reasonable interior solution for the exterior Kerr metric. Although it is possible to match numerically the Kerr solution with the interior field of an infinitely tiny rotating disk of dust \cite{meinel}, such a hypothetical system does not seem to be of relevance to describe astrophysical compact objects. 
It is now widely believed that the Kerr solution is not appropriate to describe the exterior field of rapidly rotating compact objects. Indeed, the Kerr metric takes into account the total mass and the angular momentum of the body. However, the moment of inertia is an additional characteristic of any realistic body which should be considered in order to correctly describe the gravitational field. As a consequence, the multipole moments of the field created by a rapidly rotating compact object are different from the multipole moments of the Kerr metric. For this reason a solution with arbitrary sets of multipole moments, such as the QM solution, can be used to describe the exterior field of arbitrarily rotating mass distributions. 

To completely characterize the spacetime it is necessary to find an exact solution with a set of interior multipole moments. Due to the generality 
of the exterior solution, one can expect that the corresponding interior solution can be derived by postulating an infinite series of inner 
multipoles with arbitrary functions which can be matched one by one with the exterior solution. To see if this procedure is realizable, in this work we analyze the special case of a deformed rotating body with quadrupole moment only. 
Despite this simplification, the problem of finding interior solutions 
is at present still out of reach due, in part, to the complexity of Einstein's equations with a realistic model for the inner configuration. 
We therefore limit ourselves here to the study of approximate perfect fluid solutions. 

In fact, in the case of slowly rotating compact objects it is possible to find approximate interior solutions with physically meaningful energy-momentum tensors and state equations. Because of its physical importance, in this work we will study the Hartle-Thorne (HT) 
\cite{hartle1,hartle2} interior solution which can be coupled to an approximate exterior metric. Hereafter this solution will be denoted as the HT solution.
One of the most important characteristics of this family of solutions is that the corresponding equation of state has been constructed using realistic models for the internal structure of relativistic stars. Semi-analytical and numerical generalizations of the HT metrics with more sophisticated equations of state have been proposed by different authors. A comprehensive review of these solutions is given in \cite{stergioulas}. In all these cases, however, it is assumed that the multipole moments (quadrupole and octupole) are relatively small and that the rotation is slow. We will find 
 the explicit coordinate transformation that transforms an approximation of the QM solution into the exterior HT solution and can be matched 
with an approximate solution. In this manner, we show that the QM solution satisfies the main physical conditions to be matched with 
an interior solution.

\section{Exterior solution}
\label{sec:ext}
It is well known that the main physical information about an axisymmetric stationary vacuum gravitational field can 
be encoded in the complex Ernst potential \cite{ernst}.  For a special case of the QM solution the Ernst potential reads 
\be
E= e^{-2\psi}\ \frac{x-1- (x+1)\lambda\mu + i[ y(\lambda+\mu) + \lambda-\mu]}
{x+1- (x-1)\lambda\mu + i[ y(\lambda+\mu) - \lambda+\mu] }\ ,
\label{ernst}
\ee
where the function $\psi$ can be written in terms of the Legendre polynomials $P$ and Legendre functions $Q$ as
\be
\psi = (1-\delta) Q_0 -\delta q P_2 Q_2\ ,\ P_2 = \frac{1}{2}(3y^2-1)\ ,
\ee
\be  Q_0 =\frac{1}{2}\ln\frac{x+1}{x-1}\ ,\ Q_1 = \frac{x}{2}\ln\frac{x+1}{x-1} -x\ ,
\ Q_2 = \frac{1}{4}(3x^2-1) \ln\frac{x+1}{x-1} - \frac{3}{2}x\  .
\ee
Here $\delta$ and $q$ are arbitrary real constants. Moreover, $\lambda$ and $\mu$ are functions of $x$ and $y$ 
\be
\lambda= \alpha (x^2-1)^{1-\delta} (x+y)^{2\delta -2} e^{2\delta q \beta_-}\ ,\quad
\mu = \alpha(x^2-1)^{1-\delta} (x+y)^{2\delta -2} e^{2\delta q \beta_+}\ ,
\ee
with
\be 
\beta_\pm = \frac{1}{2}\ln\frac{(x\pm y)^2}{x^2-1} +\frac32 (1-y^2\mp xy)+\frac{3}{4}[x(1-y^2) \mp y (x^2-1)]\ln\frac{x-1}{x+1}\ .
\ee
The constant $\alpha$ can be represented in terms of the additional constants $a$ and $m$ as
\be
\alpha= \frac{\sigma - m}{a} \ ,\quad \sigma = \sqrt{m^2-a^2}\ .
\ee
The solution is asymptotically flat and possesses the following independent parameters: $m$, $a$, $\delta$, and $q$. 
In the limiting case $\alpha=0$, $a=0$, $q=0$ and $\delta =1$, the only independent parameter is $m$ and the Ernst potential (\ref{ernst})
determines the Schwarzschild spacetime. Moreover, for $\alpha =a =0$ and $q=0$ we obtain the Ernst potential of the ZV static 
solution which is characterized by the parameters $m$ and $\delta$. Furthermore, for   $\alpha =a =0$ and 
$\delta =1$, the resulting solution coincides with the ER static spacetime \cite{erro59}. The Kerr metric is also contained as a special case for $q=0$ and $\delta =1$.

The physical 
significance of the parameters entering the potential (\ref{ernst})
can be established in an invariant manner by calculating the relativistic Geroch--Hansen multipole moments.
We use here the procedure formulated in \cite{quev89} which allows us to derive the gravitoelectric $M_n$ as well as the 
gravitomagnetic $J_n$ multipole moments. A lengthly but straightforward calculation yields 
\be 
M_{2k+1} = J_{2k}=0 \ ,  \quad k = 0,1,2,... 
\ee 
\be 
M_0= m + \sigma(\delta -1)
\ee
\be
M_2 = \frac{2}{15} \sigma^3 \delta q - \frac{1}{3}\sigma^3 (\delta^3 -3\delta^2-4\delta + 6) - m\sigma^2 \delta (\delta -2) - 3m^2 \sigma (\delta -1)
-m^3 \ ,
\ee
\be
J_1 = ma + 2a \sigma (\delta -1)\ ,
\ee
\be
J_3 = \frac{4}{15} a\sigma^3 \delta q - a\left[ \frac{2}{3} \sigma^3 (\delta^3 - 3\delta^2 - \delta + 3) + m\sigma^2 (3\delta^2 - 6 \delta + 2)
+ 4 m^2 \sigma (\delta -1) + m^3\right]  \ .
\ee
The even gravitomagnetic and the odd gravitoelectric multipoles vanish identically because the solution possesses and additional reflection 
symmetry with 
respect to the hyperplane $y=0$. Higher odd gravitomagnetic and even gravitoelectric multipoles can be shown to be linearly dependent since they are completely determined in terms 
of the parameters $m$, $a$, $q$ and $\delta$. From the above expressions we see that the ZV \cite{zip66,voor70} parameter 
changes the value
of the total mass $M_0$ as well as the angular momentum $J_1$ of the source. The mass quadrupole $M_2$ can be interpreted as a nonlinear superposition
of the quadrupoles corresponding to the ZV, ER and Kerr spacetimes. 
Notice that in the limiting static case of the ZV metric the only non-vanishing parameters are $m=\sigma$ and $\delta$ so that all gravitomagnetic 
multipoles vanish and we obtain $ M_0= m\delta$ and $M_2 = \delta m^3 (1-\delta^2)/3$ for the leading gravitoelectric multipoles. This means that 
the ZV solution represents the field of a static deformed body. To our knowledge, this is the simplest generalization of the Schwarschild solution 
which includes a quadrupole moment.

In order to completely describe the geometric and physical properties of the spacetime,  it is convenient to calculate the explicit form of the 
metric. In fact, the Ernst potential is defined as 
\be
E = f + i \Omega \ , \quad {\rm with}\quad \sigma (x^2-1)\Omega_x = f^2\omega_y\ , \ \sigma (1-y^2)\Omega_y = - f^2 \omega_x \ ,
\ee
where $f$ and $\omega$ are the main metric functions which determine the line element in prolate spheroidal coordinates
$(t,x,y,\varphi)$:
\be
 ds^2 = f (dt-\omega d\varphi)^2 - \frac{\sigma^2}{f}\left[ e^{2\gamma}(x^2-y^2)\left( \frac{dx^2}{x^2-1} + \frac{dy^2}{1-y^2} \right) 
+ (x^2-1)(1-y^2) d\varphi^2\right] \ .
\label{lel}
\ee
The metric function $\gamma$ can be calculated by quadratures once $f$ and $\omega$ are known. The calculation of the metric functions
results in 
\begin{equation}\label{q2}
f=\frac{R}{L}e^{-2q\delta P_{2}Q_{2}},
\end{equation}
\begin{equation}\label{q3}\omega=-2a-2\sigma\frac{{M}}{R}e^{2q\delta P_{2}Q_{2}},
\end{equation}
\begin{equation}\label{q4}
e^{2\gamma}=\frac{1}{4}\left(1+\frac{m}{\sigma}\right)^{2}\frac{R}{(x^{2}-1)^{\delta}}e^{2\delta^{2}\hat{\gamma}},
\end{equation}
where
\begin{equation}\label{q5}
R =a_{+}a_{-}+b_{+}b_{-}, \qquad   L=a_{+}^{2}+b_{+}^{2},
\end{equation}
\begin{equation}\label{q6}
{M} =(x+1)^{\delta-1}\left[x(1-y^{2})(\lambda+\eta)a_{+}+y(x^{2}-1)(1-\lambda\eta)b_{+}\right],
\end{equation}
\bea\label{q7}
\hat{\gamma}&=&\frac{1}{2}(1+q)^{2}\ln\frac{x^{2}-1}{x^{2}-y^{2}}+2q(1-P_{2})Q_{1}+q^{2}(1-P_{2})[(1+P_{2})(Q_{1}^{2}-Q_{2}^{2})\\
&&+\frac{1}{2}(x^{2}-1)(2Q_{2}^{2}-3xQ_{1}Q_{2}+3Q_{0}Q_{2}-Q_{2}^{\prime})].
\eea
Furthermore
\begin{equation}\label{q8}
a_{\pm}=(x\pm1)^{\delta-1}[x(1-\lambda\eta)\pm(1+\lambda\eta)],
\end{equation}
\begin{equation}\label{q9}
b_{\pm}=(x\pm1)^{\delta-1}[y(\lambda+\eta)\mp(\lambda-\eta)]\ .
\end{equation}

\begin{figure}
\includegraphics[scale=0.2]{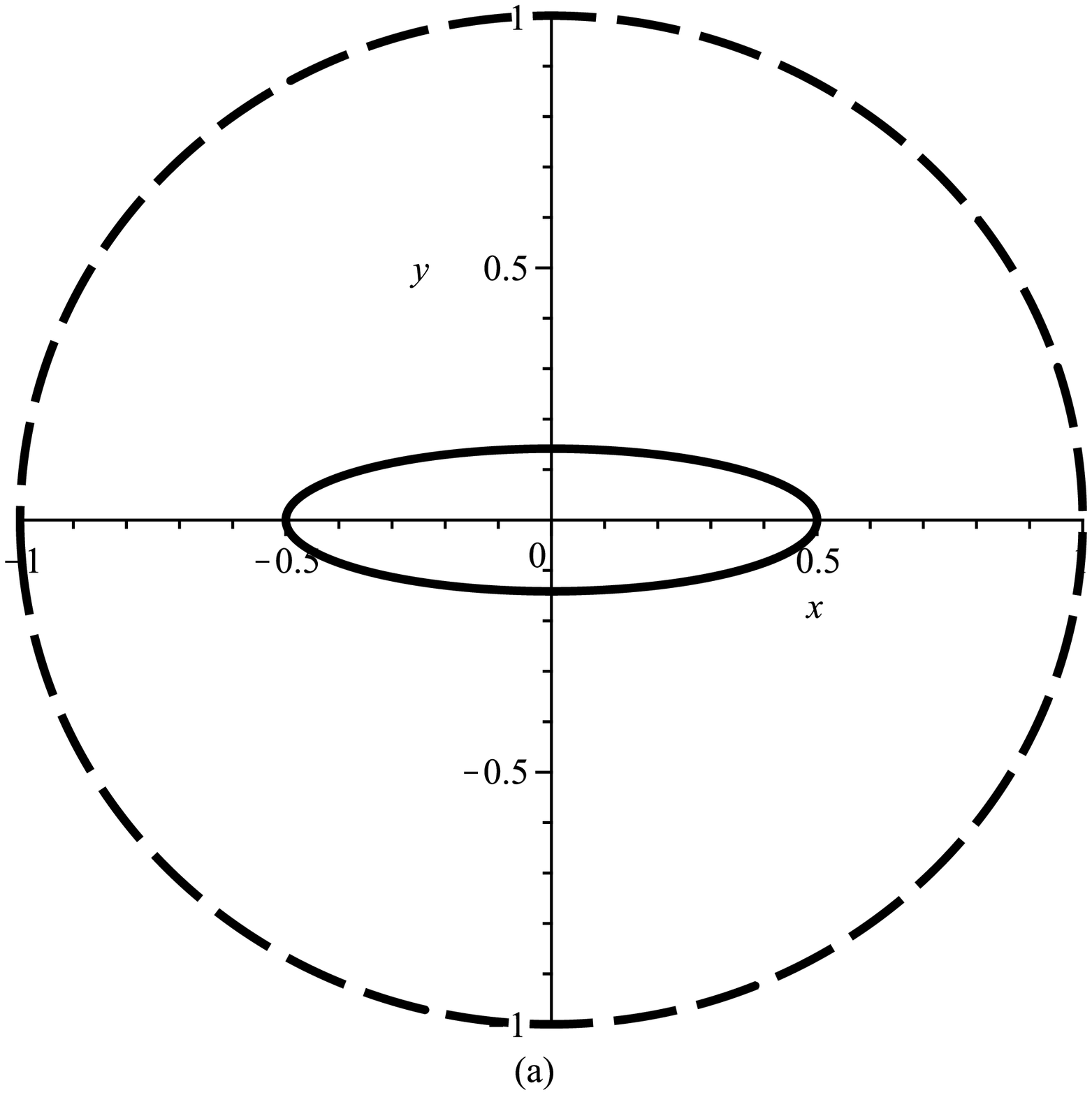}
\includegraphics[scale=0.2]{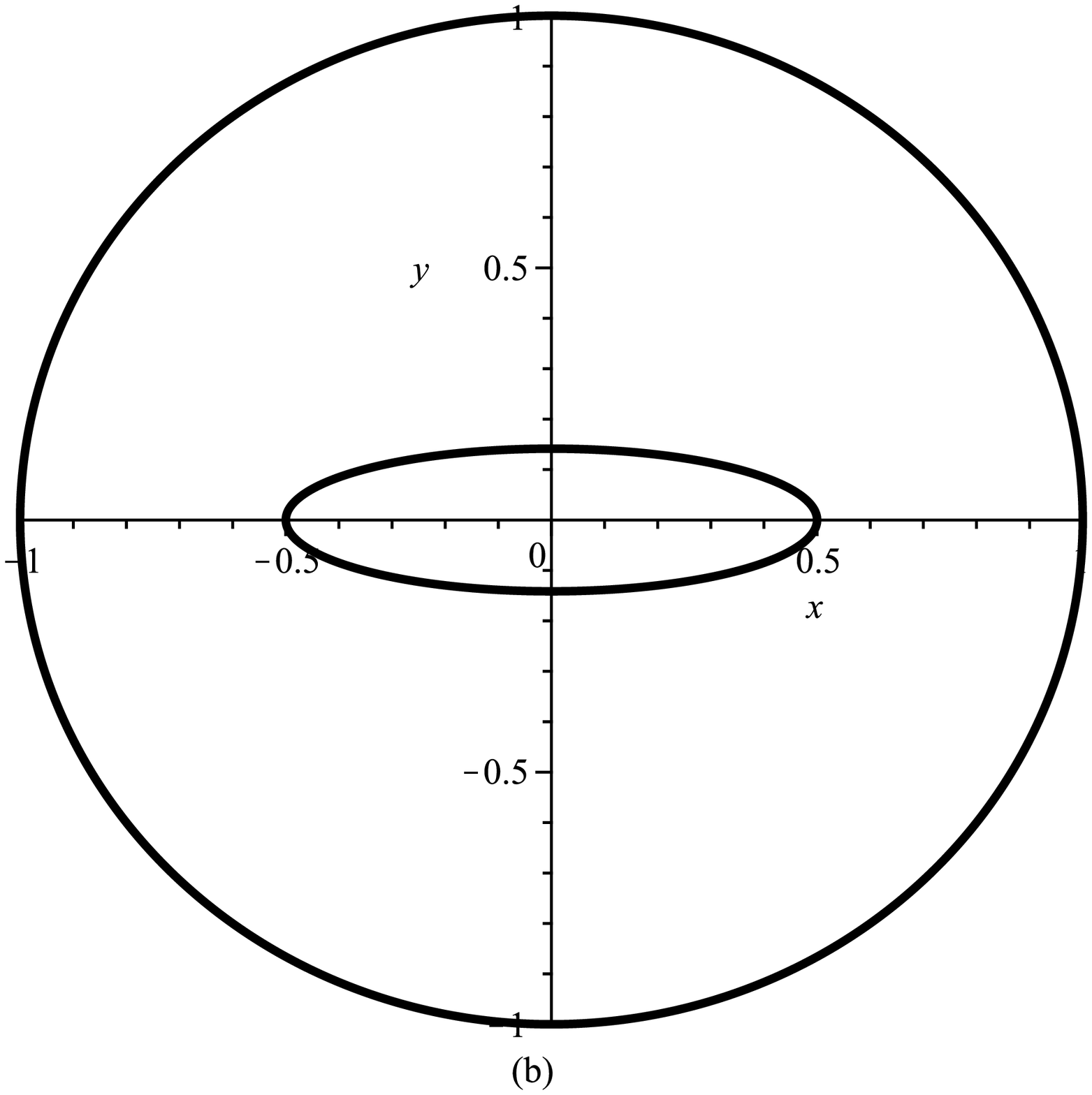}
\includegraphics[scale=0.2]{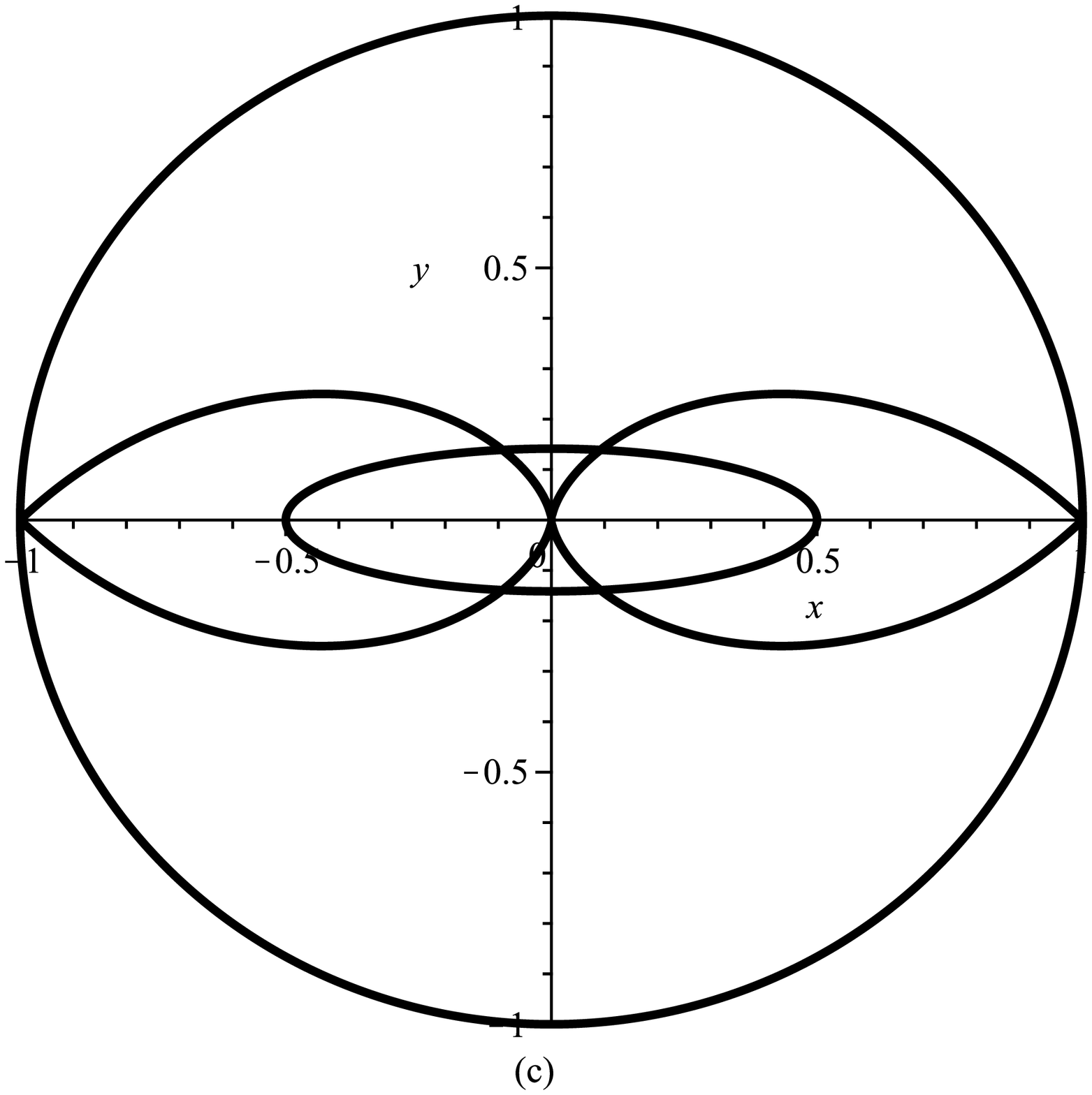}
\caption{Structure of the hypersurface $x=1$ in the spacetime of a rotating mass with arbitrary quadrupole parameter. The plot (a) 
illustrates the case of vanishing quadrupole, $q=0$, with $\delta=1$. The hypersurface $x=1$ corresponds to an event horizon (dashed line).
The horizon covers the ring singularity which is caused by the rotation of the body.   
The case (b) is for $\delta =2$ and $q=10$. The horizon $x=1$ becomes singular due to the presence of the quadrupole moment so that 
the spacetime is characterized by a naked singularity (solid line). By varying the value of the quadrupole parameter $q$ it is possible 
to generate additional singularities (c) which are always inside the outer naked singularity situated at $x=1$.  }
\label{fig1}
\end{figure}

It is easy to show that this solution is asymptotically flat and free of singularities outside the sphere $x=1$ which represents a 
naked singularity. In fact, a numerical investigation of the Kretschmann invariant shows that the hypersurface $x=1$ is singular,
independently of the value of the parameters $m$, $\delta\neq 1$, $a$, and $q$. Only in the limiting case $q=0$ and $\delta =1$, 
the hypersurface $x=1$ coincides with the exterior horizon of the Kerr spacetime. This situation is illustrated in Fig.\ref{fig1} which 
shows that a naked singularity is always present when the quadrupole parameter $q$ is non zero or when $\delta\neq 1$.
Moreover, the symmetry axis $y=\pm 1$ is free of singularities (except at $x=1$), indicating that 
the condition of elementary flatness is satisfied \cite{solutions}.

All the properties mentioned above seem to indicate that the solution can be used 
to describe the exterior field of a deformed, rotating mass distribution. From the point of view of general relativity, however, to
describe the entire manifold of a rotating body it is necessary to find an interior solution which takes into account the inner 
properties of the body and that can be matched to the exterior vacuum solution. 
However, rather few exact stationary solutions that involve a matter distribution in rotation are to be found in the literature. 
In particular, the interior solution for the rotating Kerr solution is still unknown. In fact, the quest for a realistic exact 
solution, representing both the interior and exterior gravitational field generated by a self-gravitating axisymmetric distribution 
of mass of perfect fluid in stationary rotation is considered a major problem in general relativity. We believe that the inclusion 
of a quadrupole in the exterior and in the interior solutions adds a new physical degree of freedom that could be used to search for realistic
interior solutions. To see if this is true, we will use a special limit of the above solution with quadrupole moment and will show that 
in fact it can be matched with a realistic approximate interior solution.

\section{The field of a slowly rotating and slightly deformed body}
\label{sec:approx}

In this section we calculate the limiting case where the deviation from spherical symmetry is small and the body is slowly rotating. It is 
convenient to introduce the new spatial coordinates $r$ and $\theta$ by means of \cite{bglq09}
\be
x=\frac{r-m}{\sigma} \ , \quad y =\cos\theta \ ,
\ee
and to choose the ZV parameter as $\delta=1+sq$, where $s$ is a real constant. Then, expanding the metric (\ref{lel}) to first order in the quadrupole parameter $q$ and to second order in the rotation parameter $a$, we obtain 
\bea
\label{qma1}
&f&=1-\frac{2m}{r}+\frac{2a^{2}m\cos^{2}\theta}{r^{3}}+q(1+s)\left(1-\frac{2m}{r}\right)\ln\left(1-\frac{2m}{r}\right)
 \nonumber\\
&&+3q\left(\frac{r}{2m}-1\right)\bigg[\left(1-\frac{m}{r}\right)\left(3\cos^{2}\theta-1\right) \nonumber\\
&&+ \left\{\left(\frac{r}{2m}-1\right)(3\cos^{2}\theta-1)-\frac{m}{r}\sin^{2}\theta\right\}\ln\left(1-\frac{2m}{r}\right)\bigg],
\eea
\begin{equation}
\label{qma2}
\omega=\frac{2amr\sin^{2}\theta}{r-2m},
\end{equation}
\bea
\gamma&=&\frac{1}{2}\ln\frac{r(r-2m)}{(r-m)^{2}-m^{2}\cos^{2}\theta}+\frac{a^{2}}{2}\left[\frac{m^{2}\cos^{2}\theta\sin^{2}\theta}{r(r-2m)((r-m)^{2}-m^{2}\cos^{2}\theta)}\right]\nonumber\\
&&+q(1+s)\ln\frac{r(r-2m)}{(r-m)^{2}-m^{2}\cos^{2}\theta}\nonumber\\
&&-3q\left[1+\frac{1}{2}\left(\frac{r}{m}-1\right)\ln\left(1-\frac{2m}{r}\right)\right]\sin^{2}\theta.
\label{qma3}
\eea
Furthermore, we introduce coordinates $R$ and $\Theta$ by means of   
\bea
r&=&R+{\mathcal M}q+\frac{3}{2}{\mathcal M}q\sin^{2}\Theta\left[\frac{R}{\mathcal M}-1+\frac{R^{2}}{2{\mathcal M}^{2}}\left(1-\frac{2{\mathcal M}}{R}\right)\ln\left(1-\frac{2{\mathcal M}}{R}\right)\right]\nonumber\\
&&-\frac{a^{2}}{2R}\left[\left( 1+\frac{2{\mathcal M}}{R}\right)\left( 1-\frac{{\mathcal M}}{R}\right)-\cos^{2}\Theta \left( 1-\frac{2{\mathcal M}}{R}\right)\left( 1+\frac{3{\mathcal M}}{R}\right)\right] ,
\eea
\begin{equation}
\theta=\Theta-\sin\Theta\cos\Theta\left\{\frac{3}{2}q\left[2+\left(\frac{R}{{\mathcal M}}-1\right)\ln\left(1-\frac{2{\mathcal M}}{R}\right)\right]+\frac{a^{2}}{2R}\left(1+\frac{2{\mathcal M}}{R}\right)\right\}\ ,
\end{equation}
where
\begin{equation}
{\mathcal M}=m(1-q),\ \ J=-ma,\ \ Q=\frac{J^{2}}{m}-\frac{4}{5}m^{3}q\ .
\end{equation}

A straightforward computation shows that the metric (\ref{qma1})--(\ref{qma3})  can be written as
\bea
\label{ht1}
ds^2&=&\left(1-\frac{2{\mathcal M }}{R}\right)\left[1+2k_1P_2(\cos\Theta)+2\left(1-\frac{2{\mathcal M}}{R}\right)^{-1}\frac{J^{2}}{R^{4}}(2\cos^2\Theta-1)\right]dt^2\nonumber \\
&&-\left(1-\frac{2{\mathcal M}}{R}\right)^{-1}\left[1-2k_2P_2(\cos\Theta)-2\left(1-\frac{2{\mathcal M}}{R}\right)^{-1}\frac{J^{2}}{R^4}\right]dR^2\nonumber \\
&&-R^2[1-2k_3P_2(\cos\Theta)](d\Theta^2+\sin^2\Theta d\phi^2)+4\frac{J}{R}\sin^2\Theta dt d\phi\,
\eea
where
\begin{eqnarray}\label{ht2}
k_1&=&\frac{J^{2}}{{\mathcal M}R^3}\left(1+\frac{{\mathcal M}}{R}\right)+\frac{5}{8}\frac{Q-J^{2}/{\mathcal M}}{{\mathcal M}^3}Q_2^2\left(\frac{R}{{\mathcal M}}-1\right)\ , \nonumber\\
k_2&=&k_1-\frac{6J^{2}}{R^4}\ , \nonumber\\
k_3&=&k_1+\frac{J^{2}}{R^4}-\frac{5}{4}\frac{Q-J^{2}/{\mathcal M}}{{\mathcal M}^2R}\left(1-\frac{2{\mathcal M}}{R}\right)^{-1/2}Q_2^1\left(\frac{R}{\mathcal M}-1\right)\ ,\nonumber
\end{eqnarray}
and we choose for the free parameter $s$ the particular value $s=-1$ for the sake of simplicity.
Here $Q_l^m$ are the associated Legendre functions of the second kind
\begin{equation}
Q_{2}^{1}(x)=(x^{2}-1)^{1/2}\left[\frac{3x^{2}-2}{x^{2}-1}-\frac{3}{2}x\ln\frac{x+1}{x-1}\right],
\ \ Q_{2}^{2}(x)=\frac{3}{2}(x^{2}-1)\ln\frac{x+1}{x-1}+\frac{5x-3x^{3}}{x^{2}-1}.
\nonumber
\end{equation}
The line element (\ref{ht1}) was first derived in an equivalent representation by Hartle and Thorne \cite{hartle1,hartle2}. 
Consequently, our results show that the general solution (\ref{lel})-(\ref{q9}) contains as a special case the HT 
metric which describes the exterior gravitational field of any slowly and rigidly rotating deformed body. 

In the case of ordinary stars, such as the Sun, the smallness of the parameters 
${\mathcal M_{Sun}}/{\mathcal R_{Sun}}\approx  10^{-6},$  
${ J_{Sun}}/{\mathcal R_{Sun}^{2}}\approx 10^{-12}$, 
${ Q_{Sun}}/{\mathcal R_{Sun}^{3}}\approx 10^{-12}$, 
allows us to simplify the  metric (\ref{ht1}) to obtain \cite{bkqr09}
\bea
\label{ht4}
ds^2&=&\left[1-\frac{2{\mathcal M}}{R}+\frac{2Q}{R^{3}}P_2(\cos\Theta)\right]dt^2 +\frac{4J}{R}\sin^2\Theta dt d\phi
-\left[1+\frac{2{\mathcal M }}{R}-\frac{2Q}{R^{3}}P_2(\cos\Theta)\right]dR^2
\nonumber  \\
&&
-\left[1-\frac{2Q}{R^{3}}P_2(\cos\Theta)\right]R^2(d\Theta^2+\sin^2\Theta d\phi^2)
\ .
\eea
The accuracy of this metric is of one part in $10^{12}$. Consequently, it describes the gravitational field for a wide range of compact objects, and only in the case of very dense ($ \mathcal M \sim  \mathcal R$)
or very rapidly rapidly rotating ($  J \sim \mathcal R ^ 2$) objects large discrepancies will appear. 

\section{The interior solution}
\label{sec:htint}

If a compact object is rotating slowly, the calculation of its equilibrium properties reduces drastically because it can be considered as a linear perturbation of an already-known non-rotating configuration. A very good approximation of the internal structure of the body is delivered 
by a perfect fluid model satisfying a one-parameter equation of state, $\mathcal{P}=\mathcal{P}(\mathcal{E})$, where $\mathcal{P}$ is the pressure and $\mathcal{E}$ is the density of total mass-energy. A further simplification follows from the assumption that 
the configuration is symmetric with respect to an arbitrary axis which can be taken as the rotation axis. Furthermore, the rotating object should be invariant with respect to reflections about a plane perpendicular to the axis of rotation, i.e, about the equatorial plane. 
As for the rotation, it turns out that  configurations which minimize the total mass-energy (e.g., all stable configurations) must rotate uniformly \cite{hartle3}. Uniform rotation facilitates the study of the internal properties of the body, especially if we restrict ourselves to the case of
slow rotation. That is, we assume that angular velocities $\Omega$ are small enough so that the fractional changes in pressure, energy density and gravitational field due to the rotation are all less than unity, i.e. 
$ \Omega^2 \ll \left(\frac{c}{\cal R}\right)^2 \frac{G{\cal M}}{c^2 {\cal R}}$
where ${\cal M}$ is the mass and ${\cal R}$ is the radius of the non-rotating configuration. The above condition is equivalent to the physical requirement $\Omega\ll c/{\cal R}$. The metric functions must be found by solving Einstein's equations
\begin{equation}
R_{\mu}^\nu-\frac{1}{2}\delta_{\mu}^{\nu}R ={8\pi}T_{\mu}^\nu \ ,
\end{equation}
where the stress-energy tensor is that of a perfect fluid
\begin{equation}
T_{\mu}^\nu=(\mathcal{E}+\mathcal{P})u^{\nu}u_{\mu}-\mathcal{P}\delta_{\mu}^{\nu}.
\end{equation}
The 4-velocity which satisfies the normalization condition
$u^{\mu}u_{\mu}=1$ is 
\begin{equation}
u^R=u^\Theta=0,\ \ u^\phi=\Omega u^t,\ \ u^t=\left(g_{tt}+2\Omega g_{t\phi}+\Omega^{2}g_{\phi\phi}\right)^{-1/2}\ ,
\end{equation}
where the angular velocity $\Omega$ is a constant throughout the fluid.

The explicit integration of the inner Einstein's equations depends on the particular choice of the density function $\mathcal{E}(R)$, where
$R$ is the radial coordinate, and on the equation of state $\mathcal{P}=\mathcal{P}(\mathcal{E})$. For the sake of simplicity, we consider 
here only the simplest case with $\mathcal{E}=$const and let the pressure $\mathcal{P}$ be determined by the field equations. Then, 
the total mass of the non-rotating body is ${\cal M}=4\pi  \mathcal{E} {\cal R}^3 / 3$, where ${\cal R}$ is the radius of the body. Under these
conditions, the resulting line element can be written as 
\bea
\label{h1}
ds^2 & = &\left(1+{2\Phi}{}\right)dt^2-
\left[1+{2R}{}\frac{d\Phi_{0}(R)}{dR}+\Phi_{2}(R) P_2(\cos\Theta)\right]dR^2\nonumber\\
 && -R^2\left[1+{2\Phi_{2}(R)}{}P_2(\cos\Theta)\right][d\Theta^2+\sin^2\Theta (d\phi-\tilde{\omega} dt)^2],
\eea
where
\begin{equation}
\Phi=\Phi_{0}(R)+\Phi_{2}(R)P_2(\cos\Theta),
\end{equation}
is the interior Newtonian potential. Here $\Phi_{0}$ is the interior
 Newtonian potential for the  non-rotating configuration and
$\Phi_{2}(R)$ is the perturbation due to the rotation. Moreover, 
\be
\tilde{\omega}=\Omega-\bar{\omega}
\ee 
is the angular velocity of the local
inertial frame, where $\Omega=$const is the angular velocity of the fluid 
and $\bar{\omega}=\bar\omega(R)$ is the angular
velocity of the fluid relative to the inertial frame. With a suitable choice of the 
integration constant, the interior unperturbed Newtonian potential can be
expressed as
\begin{equation}
\label{phi0}
\Phi_{0}(R)=-2\pi \rho\left({\cal R}^{2}-\frac{R^{2}}{3}\right),
\end{equation}
where $\cal R$ is the radius of the non-rotating configuration. The function
$\Phi_{2}(R)$ 
satisfies the following equation
\begin{equation}
\label{inner1}
\frac{d\Phi_{2}(R)}{dR}=\Phi_{2}(R)\left(\frac{4\pi
R^{2}\rho}{\mathcal M}-\frac{2}{R}\right)-\frac{2\chi}{{\mathcal
M}}+\frac{4\pi}{3{\mathcal M}}\rho\Omega^{2}R^{4},
\end{equation}
where $\chi$ is defined by
\begin{equation}
\label{inner2}
\frac{d\chi}{dR}=- \frac{2{\mathcal
M}}{R^{2}}\Phi_{2}(R)+\frac{8\pi}{3}\rho\Omega^{2}R^{3}\ ,
\end{equation}

For the angular velocity relative to the local inertial frame 
$\bar{\omega}$ one obtains the differential equation
\begin{equation}
\label{inner3}
\frac{1}{R^{4}}\frac{d}{dR}\left(R^4
j\frac{d\bar{\omega}}{dR}\right)+\frac{4}{R}\frac{dj}{dR}\bar{\omega}=0, \quad
j(R)=\left[1+\frac{4\pi}{3}\rho R^2\right]^{1/2} ,
\end{equation}
which is related to the equilibrium condition of a rotating self-gravitating body, namely, the 
condition that there exists   
a balance between pressure forces, gravitational forces and centrifugal forces. 
The solution for $\bar\omega(R)$ must be regular at the origin and outside the body  it must 
take the form
$\bar{\omega}(R)=\Omega-2J/R^{3}$, 
where $J$ is the total angular momentum of the star. This condition guarantees a smooth matching of the 
corresponding metric functions on the surface of the mass distribution. 

For the integration of the field 
equations it is convenient to consider also an additional differential 
equation which follows from the conservation law $T^{\mu\nu}_{\ \ ;\nu} = 0$ and involves the pressure and
its partial derivatives. Even in the simple case of $\mathcal{E}= $ const, 
the resulting expression is rather cumbersome. A detailed analysis of this equation will be presented elsewhere \cite{bkqr09}.
The integration of the system of partial differential equations (\ref{inner1})--(\ref{inner3}), together with the
differential equation for the pressure, cannot be carried out analytically. Nevertheless, it is possible to find 
numerical solutions not only in the simple case $\mathcal{E}=$ const, but also in the case of more realistic 
density functions \cite{hartle2}. 

It is easy to show analytically that the interior solution (\ref{h1}) can be matched smoothly on the surface $R=\mathcal{R}$
with the exterior solution (\ref{ht1}) in the special 
case of an unperturbed configuration with $\Phi_2 =0$ and $\Phi_0$ as given in Eq.(\ref{phi0}). In the general case 
of a rotating deformed body, the matching can be performed only numerically. In fact, the solutions of the field equations 
for the inner distribution of mass are calculated using the matching conditions as boundary conditions. In this manner, it is
possible to find realistic solutions that describe the interior and exterior gravitational field of slowly rotating
and slightly deformed mass distributions.

\section{Conclusions}
\label{sec:con}

In this work we studied an exact solution of Einstein's vacuum field equations which represents the exterior gravitational field
of a stationary axisymmetric rotating mass distribution. It contains four free parameters that determine the total mass, angular 
momentum, and quadrupole momentum of the body. It was shown that the solution is asymptotically flat and is free of singularities
outside a region that can be covered by an interior solution. In fact, we show that the presence of a gravitoelectric 
quadrupole changes drastically 
the geometric properties of the spacetime. Independently of the value of the quadrupole, there always exists  a singular surface 
which is not surrounded by an event horizon. This implies that the gravitoelectric quadrupole moment of the source can always be 
associated with the presence of naked singularities. However, the naked singularities are situated very close to the origin of
coordinates so that a physically reasonable interior solution could be used to cover them. 

The main goal of this work was to take the first step to prove that the exact exterior QM solution can be matched in general with 
an interior solution. To this end we calculated the limiting case of an approximate solution for a slowly rotating and slightly deformed
source. This means that the solution is expanded to first order in the quadrupole parameter and to second order in the rotation
parameter. We show that a particular choice of the ZV parameter allows us to find explicitly a coordinate transformation that 
transforms the solution into the approximate exterior Hartle metric.  Assuming that the inner configuration can be described by 
a perfect fluid which rotates uniformly with respect to an axis that coincides with the axis of symmetry and that the source 
is symmetric with respect to reflections about the equatorial plane, the set of differential equations which follow from Eintein's 
interior equations was derived and investigated. A particular analytic solution was presented that can be matched with the approximate 
exterior solution in the case of an unperturbed non-rotating perfect fluid. To consider the general case of a slowly rotating and 
slightly deformed perfect fluid it is necessary to perform numerical calculations to integrate the corresponding set of differential 
equations. Our numerical results show that it is possible to obtain numerical solutions with realist physical properties and to 
match them with the exterior approximate solution. The inclusion of the quadrupole parameter in the interior and 
exterior metrics is equivalent to adding a new  degree of freedom which facilitates to search for interior solutions.

We conclude that the particular QM solution presented in this work can be
used to describe the gravitational field of a rotating body with arbitrary quadrupole moment. In this work, it was shown that
in the case of a slightly deformed body with uniform and slow rotation, there exist a set of solutions of Einstein solutions which 
can be used to completely describe the spacetime. The case of an arbitrarily rotating perfect fluid with arbitrary quadrupole parameter will
be investigated in a future work.

\section*{Acknowledgements} 
I would like to thank K. Boshkayev, D. Bini, A. Geralico, R. Kerr and R. Ruffini for helpful comments.
I also thank ICRANet for support. 


\end{document}